# Chemistry with Graphene and Graphene Oxide - Challenges for Synthetic Chemists

Siegfried Eigler* and Andreas Hirsch*

Dr. Siegfried Eigler*, Prof. Dr. Andreas Hirsch* Department of Chemistry and Pharmacy and Institute of Advanced Materials and Processes (ZMP) Henkestrasse 42, 91054 Erlangen and Dr.-Mack Strasse 81, 90762 Fürth (Germany)
Fax: (+49) 911-6507865015
E-mail: siegfried.eigler@fau.de; andreas.hirsch@fau.de

The chemical production of graphene as well as its controlled wet- chemical modification is a challenge for synthetic chemists and the characterization of reaction products requires sophisticated analytic methods. In this review we first describe the structure of graphene and graphene oxide. We then outline the most important synthetic methods which are used for the production of these carbon based nanomaterials. We summarize the state-of-the-art for their chemical functionalization by non-covalent and covalent approaches. We put special emphasis on the differentiation of the terms graphite, graphene, graphite oxide and graphene oxide. An improved fundamental knowledge about the structure and the chemical properties of graphene and graphene oxide is an important prerequisite for the development of practical applications.

## 1. Introduction: Graphene and Graphene Oxide – Opportunities and Challenges for Synthetic Chemists

Research into graphene and graphene oxide (GO) represents an emerging field of interdisciplinary science spanning a variety of disciplines including chemistry, physics, materials science, device fabrication and nanotechnology.[1] At the same time the field of graphene and GO has a quite long lasting history.[1d-h, 2] The current graphene boom started in 2004 when Geim and Novoselov published the deposition and characterization of single sheets of graphite on solid supports.[1a] Their groundbreaking experiments on graphene were honored with the Nobel Prize in Physics in 2010.[1b, 1c, 3] Exceptional electronic, optical and mechanical properties were discovered in quick succession as a consequence of the experience gained from other carbon allotropes.[4] Especially the high charge carrier mobilities, the electrical and thermal conductivity combined with transparency and mechanical strength make graphene highly attractive for future high-tech applications.

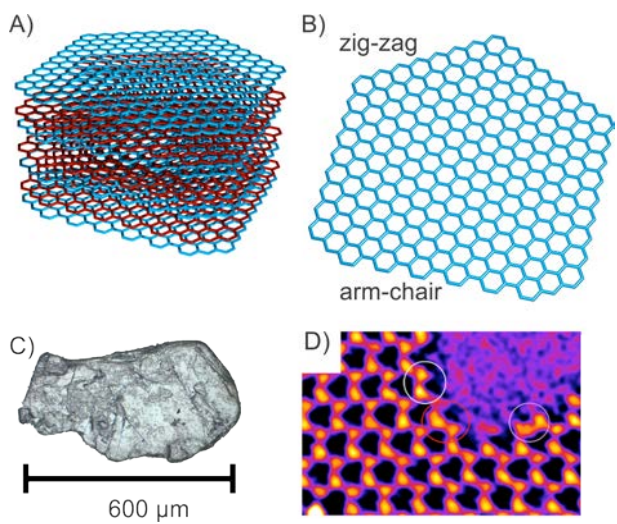

Figure 1. Schematic representation of A) the ideal structure of AB stacked graphite; B) the structure of a sheet of graphene with zig-zag and arm-chair edges; C) photography of natural graphite with visible macroscopic cracks and holes; D) HRTEM image of graphene with one edge. Adapted by permission from Macmillan Publishers Ltd: Nature Communications,[5] copyright (2014).

The current state of graphene technology with respect to prototype applications has been extensively reviewed.[6] Many graphene based devices outperform reference systems, for example in high- frequency transistors, foldable and stretchable electronic or photodetectors,[6b, 7] capacitors,[8] transparent electrodes,[9] sensors,[10] $H_2$-generation,[11] pollution management,[12] energy applications,[13] biomedical applications,[7b, 14] or in composite materials.[15]

Which role can synthetic and in particular wet chemistry play in the field of graphene- and GO-technology and can it push the field a significant step further ahead? The last 20 years have already witnessed that chemical functionalization of other synthetic carbon allotropes such as fullerenes and carbon nanotubes has led to many important accomplishments such as improvement of solubility and processability, combination of properties with other compound classes and last but not least discovery of unprecedented reactivity principles.[16] Numerous well defined covalent and non-covalent derivatives of fullerenes and nanotubes have been synthesized and many of those show outstanding properties. Conceptually, it can be expected that the chemical behavior of graphene and GO resembles those of fullerenes and carbon nanotubes, especially concerning addition reactions to the conjugated π-system. But there are also significant differences to be expected especially because in contrast to fullerenes and carbon nanotubes graphene is a flat and strain free system whose plane can be attacked from both sides when dispersed in a solvent.

Graphene is a 2D-carbon allotrope which can be viewed as both a solid and a macromolecule with molecular weights of more than $10^6$-$10^7$ g/mol. In natural graphite the graphene layers stick together by very pronounced π-π-stacking interactions. This non-covalent interlayer binding contributes significantly to the high thermodynamic stability of graphite. As a consequence wet chemistry of graphene is always accompanied with overcoming these interactions. For example, a targeted exfoliation of graphite or the stabilization of solvent-dispersed graphene sheets always competes with re-aggregation. It should be pointed out that a solid sample of graphene can only be stabilized on a support such as a surface. A non-supported graphene powder ***can not be expected to exist*** since at least partial re-stacking to graphite will take place! Another possibility of stabilizing individualized graphene is to "mask" the surface in terms of chemical functionalization.[17] Until now, it has not been demonstrated that a graphite crystal was completely solvent-dispersed into individualized graphene sheets. Although the dispersion of a certain fraction can be accomplished, assisted for example by a surfactant. Transformation of graphene into a derivative such as GO, however, can allow efficient wet chemical dispersion.[18] Despite such inherent difficulties and limitations of the wet chemistry, functionalization of graphene is a very challenging but promising approach. Many exciting hybrid systems involving covalently bound functional building blocks can be imagined. For example, combining the electrical conductivity of graphene with selective recognition sites of addends may enable *in vivo* monitoring of biomolecules. To reach such ambitious goals, it is necessary to prove the formation and stability of chemical bonds to attached addends beyond any doubt. Furthermore, it is inevitable to identify the degree of functionalization on graphene derivatives and to control possible side reactions. Within the process of the chemical synthesis of a new compound the purification and the unambiguous structural characterization of the reaction product represent key endeavors. However, in the case of graphene chemistry this is a very difficult objective because of the polydispersity, polyfunctionality and in many cases unfavorable solubility of the prepared derivatives. Moreover, classical methods used by synthetic chemists for decades to isolate and characterize new molecules such as chromatography, mass spectrometry or NMR-spectroscopy cannot be applied. Therefore, in addition to the development of successful concepts for the wet chemical functionalization of graphene and GO new analytical tools for a satisfied structure characterization have to be elaborated and applied. In this review we provide an overview on the state-of-art of the wet chemistry of graphene and GO. We will first line out inherent and important characteristics of their structural composition. Then we will discuss suitable preparation methods to make graphene and GO available as a starting material for chemical modifications. Finally, we present functionalization concepts and also discuss open challenges for the synthetic carbon allotrope chemistry.

## 2. Structure Definitions and Chemistry Concepts

In the literature one can often find the terms graphite, graphene, graphite oxide and graphene oxide used without much care and they are often intermingled, which can be misleading. For this reason we want to clarify these terms first before we start with the subsequent discussion of the wet chemistry of graphene and GO.

### 2.1. Graphite

Graphite can be of natural origin or synthetically generated.[19] The 3D stacking of the individual $sp^2$-layers can either lead to a hexagonal (AB) or rhombohedral (ABC) stacking or the structure can be turbostratic with no regularities within the layer sequence.[20] Samples of natural graphite comprise several portions of these structures that influence the reactivity.[19] The ideal structure of graphite is shown in **Figure 1**, but flakes of natural graphite usually bear macroscopic cracks and holes that can significantly determine the chemical reactivity.

### 2.2. Graphene

As depicted in **Figure 1B** graphene is a *single layer of graphite* and is built of $sp^2$-hybridized carbon atoms arranged in a honeycomb lattice. Here we use the expression $G_1$ as descriptor for graphene. The subscript "1" denotes that *exactly one layer of graphite* is considered. Accordingly two π-π stacked graphite layers are denoted as $G_2$ and can also be called bilayer graphene. An aggregate consisting of less than ten layers ($G_{<10}$) is called few-layer graphene.

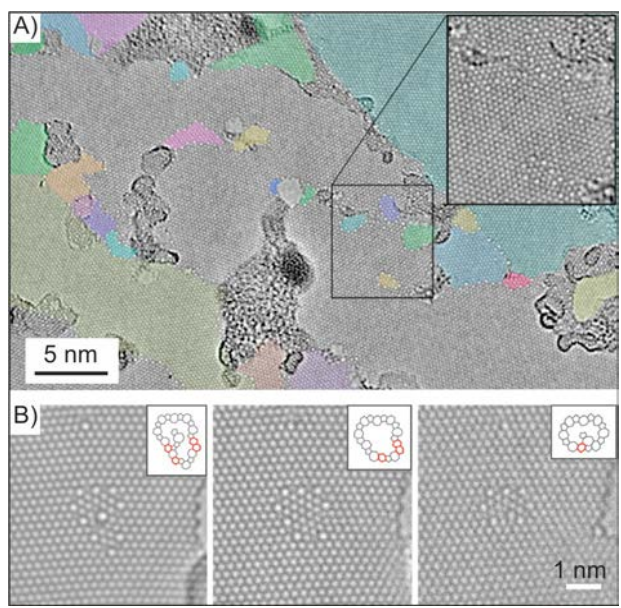

*Figure 2.* A) HRTEM image of graphene with grains marked in various colors; inset: magnification with grain boundaries; B) magnification of defect structures with five and seven membered carbon rings. Reproduced with permission.[21] Copyright 2012, American Chemical Society.

While ideal graphene would be a plane of infinite dimensions, real graphene exhibits edges that are either a zig-zag or an arm-chair arrangement. The high resolution transmission electron microscopy (HRTEM) image of graphene in **Figure 1D** shows a graphene layer with a typical edge.[5] The atomic structure of edges becomes visible. Beside edge structures lattice defects have also been studied by HRTEM.[21] The structure of graphene grown by chemical vapor deposition (CVD) on copper and subsequently transferred for analysis is shown in **Figure 2**.[21] Beside some holes, merged graphene domains and so-called grain boundaries with broken hexagonal symmetry due to five and seven membered carbon rings were found.[21-22] Structural defects can result in locally curved structures that cause local doping and therefore influence the reactivity of graphene.[23]

### 2.3. *Graphite Oxide and Graphene Oxide*

Studies on properties and applications of GO have been extensively reviewed.[6a, 24] GO is a single layer of graphite oxide. During the formation of graphite oxide the graphene layers in graphite become intercalated by an acid to form a stage 1 intercalation compound with all layers being intercalated. Subsequent oxygenation of such stage 1 intercalation compounds occurs on both sides of the basal plane and in this way graphite oxide is formed. Delamination of single layers of graphite oxide leads to GO (**Figure 3**). The exact nature of the functional groups in GO strongly depends on the reaction conditions, such as preparation time and temperature as well as on the work-up procedure. Typically GO consists of about 45 mass-% of carbon. Although several structure models have been proposed GO represents a rather polydisperse material, whose exact structure is very difficult to be precisely displayed.

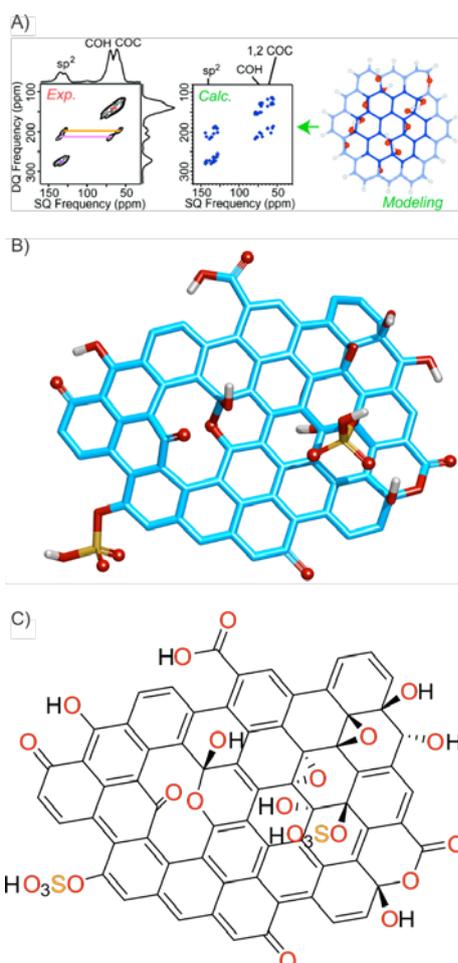

**Figure 3.** A) $^{13}C$ NMR investigation of $^{13}C$ labeled graphite oxide; Reproduced with permission.[25] Copyright 2010, American Chemical Society. B) structure model of GO with organosulfate groups in addition to hydroxyl and epoxy groups on both sides of the basal plane and hydroxyl, carbonyl and lactol groups as well as carboxylic acids at the edges. A proposed defect hole structure stabilized by a adjacent carbonyl group and a hemi-acetal due to the loss of one carbon atom is shown;[26] C) structural formula of the structure model displayed in B).

Furthermore, defects within the σ-framework of the C-skeleton can easily form upon over-oxidation. This process is always accompanied with the release of $CO_2$. These defects in GO are difficult to characterize precisely and are impossible to heal without completely reassembling the carbon framework what would require temperatures > 1500 °C.[27] As we will outline below, the defect density can be estimated after chemical reduction.[28] We have recently invented a new synthesis protocol for GO that preserves the carbon framework to a very large extend and only a minor amount of σ-hole defects are generated with a residual defect density as low as about 0.01%.[29] Therefore, this new type of GO exhibits an almost perfect honeycomb lattice. We denote this graphene oxide with an almost intact carbon framework as ai-GO. The difference between graphene derived from conventionally prepared GO and ai- GO is illustrated in **Figure 4**.

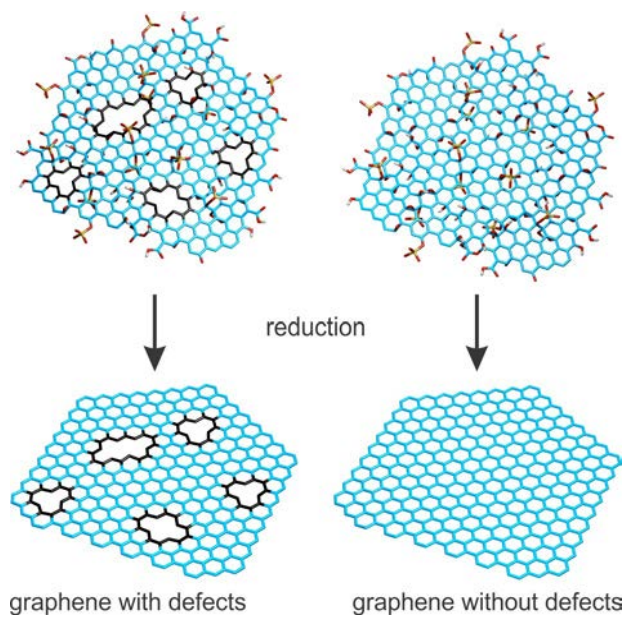

***Figure 4.*** GO with σ-hole defects can only be converted to graphene with σ-hole defects; residual functional groups at edges are omitted. On the other hand ai-GO can be reduced to almost intact graphene.[27]

### 2.3.1. "Oxo"-functionalities on graphene – and GO

The most suitable structure model for GO is based on the investigations of Lerf and Klinowski and was confirmed and advanced by the work of Ishii and Gao, respectively.[25, 26b, 30] Along these lines also $^{13}$C labeled graphite oxide was synthesized and analyzed by solid state NMR spectroscopy.[25] The results strongly suggest that hydroxyl and epoxy groups are in close proximity to each other (**Figure 3A**) and a large portion of sp$^2$-carbon remains preserved during oxidation. If GO is synthesized in sulfuric acid, GO with a sulfur content of up to 6% can be found that originates from covalently bound sulfate (**Figure 3B**).[26a] This organosulfate is hydrolytically stable in pure water at ambient conditions and can be distinguished from adsorbed inorganic sulfate. Furthermore, it contributes to the acidity of GO and enables chemical reactions.[31]

### 2.3.2. Addends at edges and defect sites

Based on NMR spectroscopy lactol groups were identified at the edges of graphene and are represented in the GO model of **Figure 3**.[26b] Other O-functionalities are carboxylic acids, hydroxyl and carbonyl groups. It should be kept in mind that the edges of graphene/graphite are either arm-chair or zig-zag. Edge oxidation leads to carbonyl or hydroxyl groups. The formation of carboxyl or lactol groups requires breaking of C-C bonds that may be accompanied with the loss of carbon induced by over-oxidation and $CO_2$ formation during synthesis, as outlined below.

Generally, following the preparation procedures by Brodie,[32] Staudenmeier[33] or Hummers[34] the loss of carbon and formation of $CO_2$ cannot be prevented. Recent results suggest that about one $CO_2$ molecule per 35-55 lattice carbon atoms is already formed during the oxidation process and the final product bears about one carbonyl group per 10-12 lattice carbon atoms.[35] The loss of carbon from the carbon-framework results consequently in permanent defects including holes of various sizes (**Figure 4**). Edges at defect sides are terminated by oxygen functionalities as indicated by a proposed structure in **Figure 3B** and **C**, respectively.

The heterogeneous structure of GO can be visualized by HRTEM imaging (**Figure 5**).[36] It comprises oxidized regions beneath small preserved aromatic regions. However, it remains difficult to visualize defects consisting of single atoms, only.[36] Further insight was provided by STM investigations on GO.[37]

These structural insights demonstrate that GO is not a defined material and it is important to keep in mind that the chemical composition, type and amount of oxygen-addends depends on the preparation procedure.

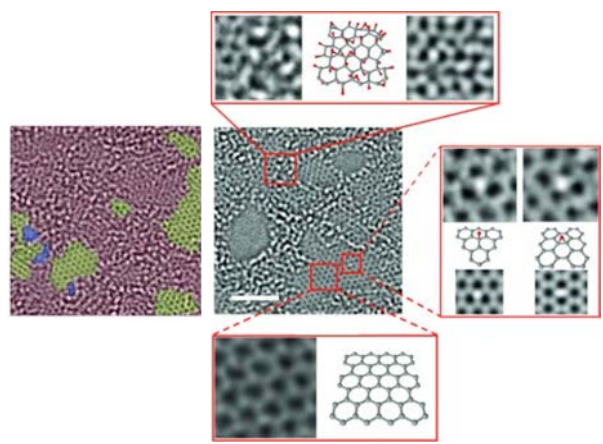

*Figure 5.* HRTEM image of GO that displays preserved regions (green) of graphene (1-2 nm), holes (blue) and heavily oxidized regions (red), insets: measured and simulated images compared with structure models. Reproduced with permission.[36] Copyright 2010, Wiley-VCH Verlag GmbH & Co.

## 3. Formation of Graphene and Graphene Oxide

Graphene generation using wet chemical approaches was accomplished by a variety of methods each having advantages and limitations. Non wet chemical methods, which provide access to small amounts of high quality graphene on surfaces, are also outlined briefly for comparison.

### 3.1. Non Wet chemical Methods for the Production of Graphene on Surfaces

A typical non wet chemical production method for graphene is based on chemical vapor deposition (CVD) on metal surfaces at about 1000 °C. A preferred surface for the synthesis is copper which can be used to make continuous films of graphene, however, with grain boundaries and the need to transfer graphene onto the desired surface.[38]

Few individual flakes of graphene can be obtained by mechanical cleavage using an adhesive tape.[1a] These flakes are visible if placed on a Si-wafer with 300 nm thick coverage of $SiO_2$ using an optical microscope, or even by the eye, what is beneficial for many investigations on single sheets of graphene.[1a, 39] Moreover, graphene from SiC (epitaxial growth) can be obtained; however, isolation of graphene remains a complex procedure.[40] These methods are not suitable for chemical bulk functionalization. Nevertheless, since the chemical structure bears very low defect densities of approximately 0.01% - 0.001% this graphene is suitable for the evaluation of reactions because reactions can be easily identified by Raman spectroscopy as explained in chapter 3.2.3.[41]

### 3.2. Wet Chemical Synthesis of Graphene Oxide and Graphene

The oxidation of graphite to graphite oxide synonymously also termed as "graphitic acid" was first described by Schafhaeutl in 1840.[42] In 1855 Brodie discovered the formation of yellow graphitic acid after oxidizing graphite in nitric acid with potassium chlorate as oxidant.[43] Staudenmaier optimized the procedure to minimize the risk of explosions caused by the accumulation of $ClO_2$.[33, 44] In 1909 Charpy described the oxidation of graphite in sulfuric acid using potassium permanganate as oxidant, keeping the temperature below 45 °C in order to suppress the extensive formation of $CO_2$ to a certain degree.[45] The same procedure, which was shown to be scalable, was later called Hummers' method.[34, 46] Hummers' procedure can be applied on a multi-gram scale in the laboratory and is the most frequently used method to prepare graphite oxide and its single layers, obtained after delamination in a suitable solvent. These single layers are called graphene oxide.

### 3.2.1. Reaction Intermediates during the Oxidation of Graphite in Sulfuric Acid with Potassium Permanganate as Oxidant

The oxidation mechanism of graphite in sulfuric acid is not fully understood. However, there is evidence for several key- intermediates. Generally, natural graphite is used as starting material to enable large-scale synthesis of GO (**Figure 6**). In the first step graphite is dispersed in sulfuric acid and becomes intercalated by sulfuric acid in the presence of an oxidant. This leads to the formation of graphite sulfate, a graphite intercalation compound (GIC).[47] The intercalation is accompanied with an increase of the layer distance resulting in an activation of graphite. It was assumed that either permanganate or *in situ* formed dimanganese heptoxide are the active oxidants.[24c] These species must be readily able to diffuse through the interlayer space of graphite sulfate. As a consequence, manganese esters are formed. It is desirable to control this process in order to prevent over-oxidation, formation of $CO_2$ and the resulting impossible-to-heal hole defects in the graphene lattice. The hydrolysis of manganese esters and the solubilization of manganese oxo-species are accomplished by the addition of water and hydrogen peroxide. It is reasonable to assume that cyclic organosulfate groups are formed during the oxidation after partial hydrolysis of manganese esters in sulfuric acid.[35] The subsequent work-up procedure either favors the hydrolysis of cyclic organosulfate to organosulfate or the complete hydrolysis that may be promoted by the action of hydrochloric acid at elevated temperatures.[26a, 48] The purification of graphite oxide is achieved by centrifugation and re-dispersion in water or by dialysis.[49] Delamination of graphite oxide to GO in water can be facilitated by sonication. GO is dispersible in water and polar solvents and can be processed as single layers by various techniques including the Langmuir-Blodgett method or by spin-coating (**Figure 6B, C**).[18, 50] The size of deposited GO flakes typically varies between few 10- 100 nm and up to 100 μm.[51]

It turned out that controlling the reaction temperature (< 5-10 °C) during both the oxidation step and especially the work-up prevents to a very large extend the over-oxidation of graphene layers. This procedure enables the isolation of GO with an almost intact σ-framework of C-atoms (ai-GO) with a defect density as low as 0.01%.[29, 52]

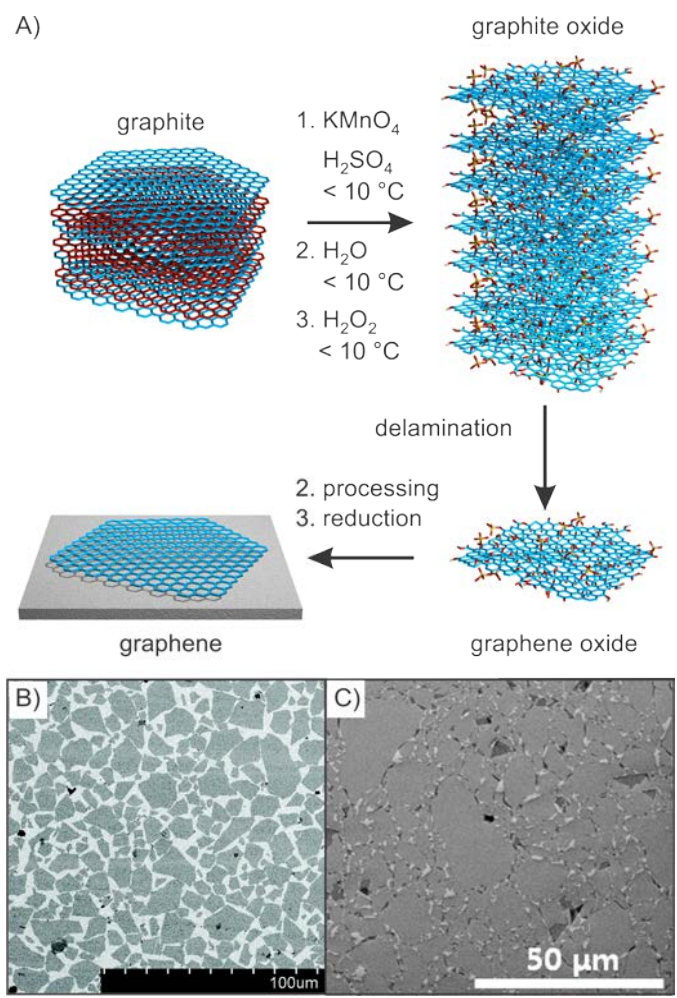

***Figure 6.*** A) Synthesis of ai-GO and graphene, starting from graphite in sulfuric acid with potassium permanganate as oxidant; SEM images of B) a Langmuir-Blodgett film of GO; [50a, 50b] Reproduced with permission.[50a] Copyright 2009, American Chemical Society; and C) a spin-coated GO film. Reproduced with permission.[50c] Copyright 2013, American Chemical Society.

### 3.2.2. Reduction of GO to Graphene

The reduction of GO to graphene has been approached with a variety of methods.[53] The most simple way is thermal annealing causing disproportion of GO into $CO_2$ and graphene. Although this method is attractive due to its simplicity, perfect graphene was not obtained, even at temperatures up to 1100 °C. Instead a ruptured carbon framework is obtained bearing σ-hole defects functionalized with oxygen functionalities such as carbonyl groups or ethers (**Figure 4**).[54] Temperatures higher than 1500 °C are required for the complete deoxygenation of GO what causes reorganization of the carbon framework. [27] Such conditions are not favorable due to high energy cost or the incompatibility with temperature sensitive substrates. Furthermore, CVD methods are superior in generating graphene at even lower temperatures with a better quality. The only reversible addition and thermal removal of oxygen atoms to graphene was reported for low concentrations of oxygen atoms in vacuum.[55] Otherwise the irreversible generation of defects within the σ-framework of C-atoms occurs. Even the attempt to repair defects within the carbon framework using small organic molecules at > 800 °C was only partially successful.[56]

Therefore, the usage of reducing agents in combination with an annealing step up to 200 °C has been targeted. Typical reducing agents are hydrazine and hydriodic acid, respectively.[53c] All methods have in common that intact graphene cannot be obtained from defective GO.

The evaluation of the local graphene domains was possible by HRTEM after reduction of GO at 800 °C using hydrogen plasma (**Figure 7**). Despite these harsh and non wet-chemical reduction conditions the intact graphene domains are not larger than 1-9 nm$^2$ at best.[57] With hydrazine as reducing agent nitrogen was found to be incorporated into the carbon lattice as revealed by NMR.[58] Scanning tunneling microscopy (STM) imaging suggests that residual defects are often decorated with oxygen functionalities, as carbonyl groups.[59]

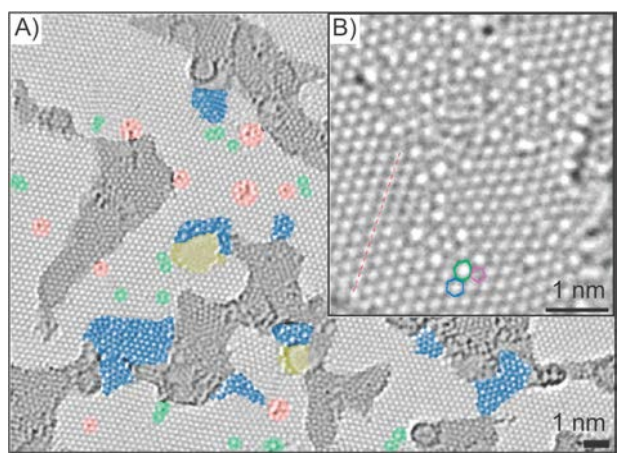

*Figure 7.* A) HRTEM image of reduced GO (reduced at 800 °C, H2) that displays preserved regions of graphene (grey), contaminated regions (dark grey); disordered regions (blue), individual ad-atoms or substituted atoms, beneath isolated topological defects (green) and holes (yellow); scale bar: 1 nm; B) magnification of a defect rich region. Reproduced with permission.[57] Copyright 2010, American Chemical Society.

As indicated above we have recently developed a methodology for the synthesis of ai-GO with an almost intact σ-framework, by preventing the evolution of $CO_2$ during synthesis by temperature control (< 5-10 °C) during oxidation and aqueous work-up.[29] The reduction of ai-GO with HI leads indeed to the formation of graphene with a defect density of about 0.01% (average distance of defects ($L_D$) up to 14 nm).[29] In this way films of graphene flakes with an average defect density of 0.08% could be produced.[52] The evaluation of $L_D$ and the defect density, respectively, can be accomplished by statistical Raman microscopy (SRM), a method that we introduced recently.[60] The efficiency of the applied reducing agents for graphene oxide was studied and it was revealed that reduction with HI is more effective than that with hydrazine or thermal treatment.[52]

*3.2.3 Determination of the defect density and the degree of functionalization by Raman spectroscopy*

Raman spectroscopy is one of the most powerful tools for the characterization of graphene, GO and their covalent derivatives.[61] The evaluation of the full-width at half-maximum (Γ) of peaks in Raman spectra can be correlated with the density of defects introduced by covalent functionalization.[62] As depicted in **Figure 8** Raman spectra display tree major peaks, the G peak, the defect activated D peak and the 2D peak. When introducing $sp^3$-defects into the basal plane of graphene all peaks broaden and the $I_D/I_G$ ratio increases to about 4 using a green laser for excitation (**Figure 8B, 9A**). At this maximum $L_D$ is about 3 nm and the defect density is about 0.3%. For $L_D$ < 3 nm the $I_D/I_G$ ratio decreases again and additional peak broadening takes place (**Figure 8C**). An idealized illustration for the degree of functionalization and defect density, respectively, with $L_D$ = 10 nm (0.03%) is represented in **Figure 8C**. Scanning films of graphene with a certain increment (µm-scale) and recording several thousands of spatially resolved spectra is the basis of statistical Raman microscopy (SRM), a very powerful analysis tool that we established recently (**Figure 9**) for the visualization of the heterogeneity of the samples.[60]

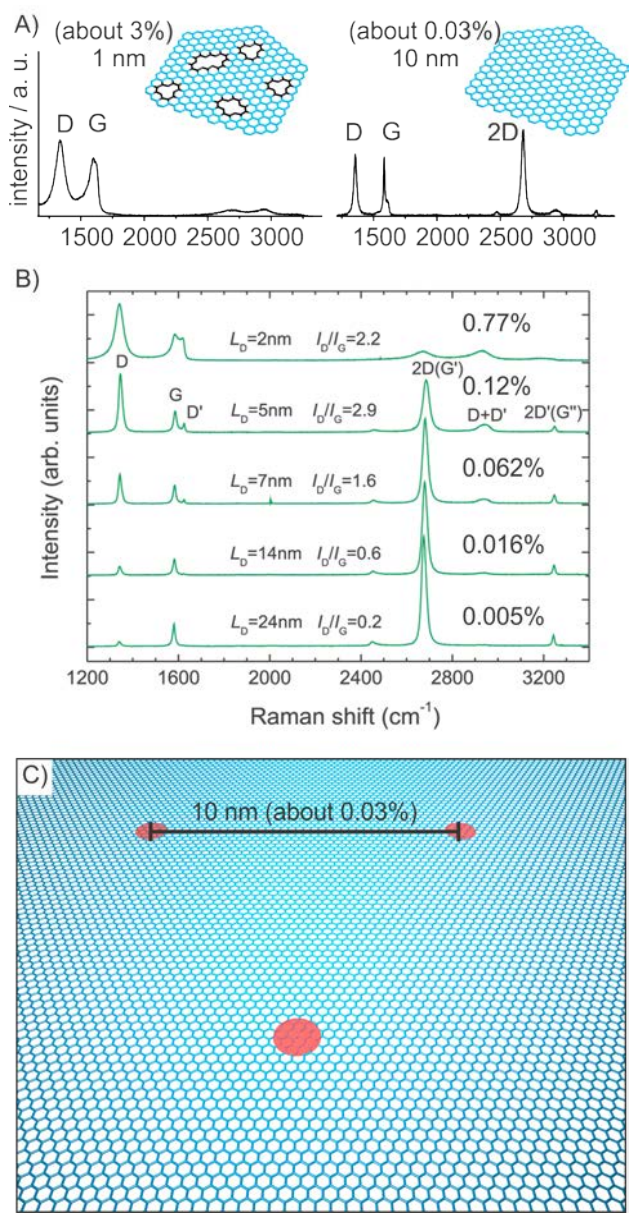

*Figure 8.* A) Raman spectra of graphene from left: GO with a defect density of 1-3% and right: ai-GO with 0.03% defects, insets: simplified structure models of graphene with defects and without defects, respectively; B) Raman spectra of graphene with LD between 2 nm and 24 nm (defect densities given in %). Reproduced with permission.[62a] Copyright 2011, American Chemical Society; C) Illustration of an idealized distance pattern of defects of 10 nm.

Raman spectroscopy in particular provides information about the integrity of the carbon framework. The intensity of the D peak increases upon successive introduction of either holes or $sp^3$-centers due to covalent addend binding. It is not possible, however, to distinguish between holes and $sp^3$-defects by Raman spectroscopy.[28, 60a] As a consequence the D-peak signal can be used for both the determination of the quality of graphene obtained by reduction of GO,[28, 52, 60b] and for the degree of functionalization of graphene.[60a] This correlation holds especially for the case where the defect density is not higher than about 1%.

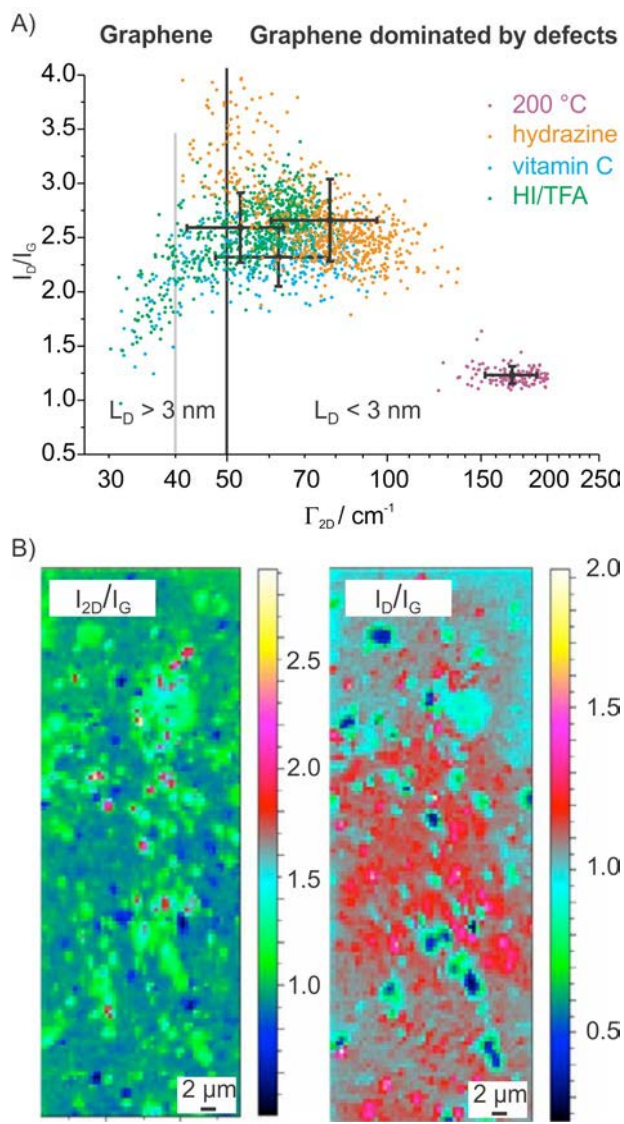

*Figure 9.* A) Illustration of Statistical Raman microscopic (SRM) analysis of films of ai-GO by plotting ID/IG *vs.* Γ2D: reduced by thermal treatment, hydrazine, vitamin C or HI/TFA ;[52] - Published by The Royal Society of Chemistry; B) SRM images of functionalized graphene from the reaction of C8K and 4-*tert*-butylphenyldiazonium tetrafluoroborate displaying local variations in films (I2D/IG and ID/IG); Adapted by permission from Macmillan Publishers Ltd: Nature Chemistry,[63] copyright (2011).

## 3.2.4 Approaches towards generation of graphene

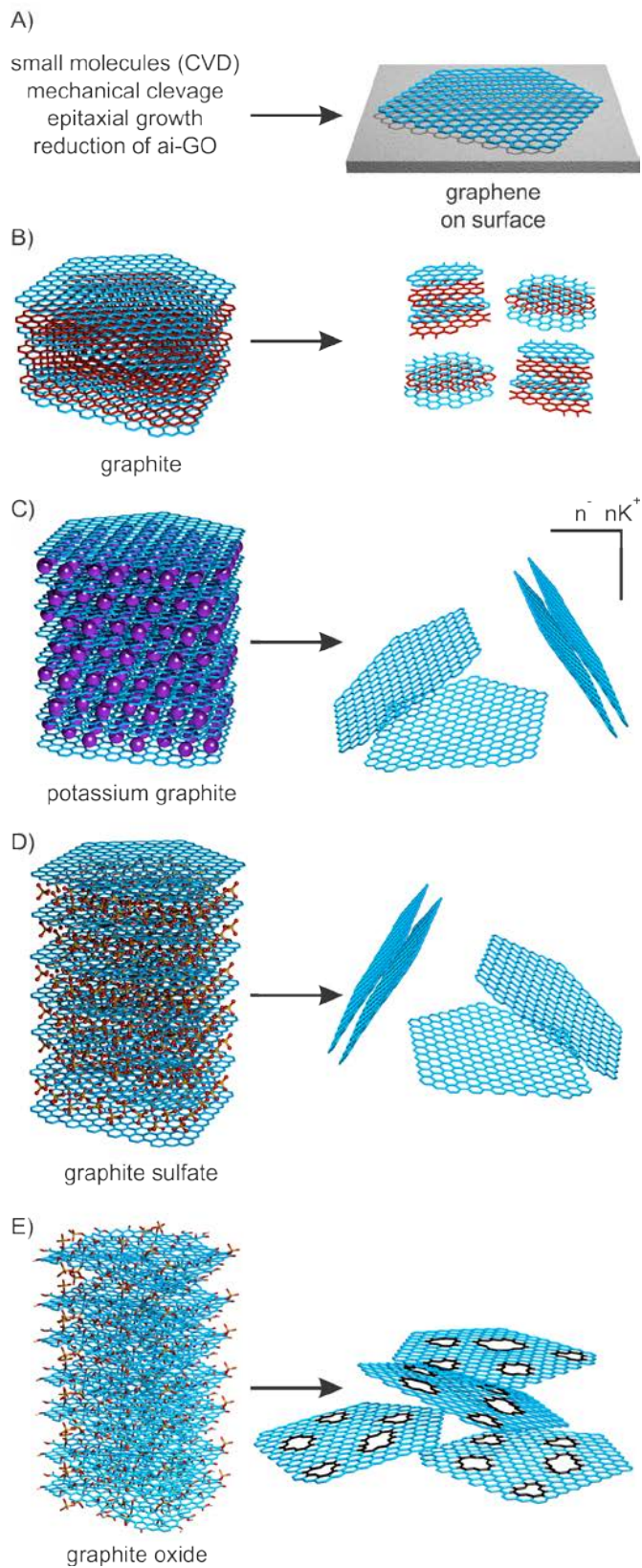

**Figure 10.** Synthetic approaches towards graphene and few-layer graphene: A) from small molecules by CVD, mechanical cleavage, epitaxial growth or from ai-GO B) from graphite by sonication in solvents or ball milling eventually with the aid of surfactants C) from donor-GICs, such as C8K in inert solvents D) from acceptor-GICs by thermal treatment or liquid exfoliation; E) from graphite oxide by thermal treatment.

The most important methods to synthesize graphene are summarized in **Figure 10**. Sheets of graphene prepared on a surface are mostly obtained either by CVD methods,[38c, 38d, 64] epitaxial growth,[65] mechanical cleavage[39] or from ai-GO.[29] The wet chemical dispersion and exfoliation of graphite was expected to be a rather attractive method for the bulk production of graphene.[66] However, despite many approaches using surfactants, e. g. sodium cholate (**Figure 10B**, **11**) in water or solvents with high boiling points, like N-methylpyrrolidone, it remains challenging to reach a quantitative stabilization of individual graphene sheets.[67] Furthermore, species adsorbed on graphene, also solvents with high boiling-points, are difficult to remove.[68]

During these exfoliation approaches few-layer graphene with a flake diameter of about 150 nm in average is formed in quite large portions. This is also due to the fact that graphite tends to break apart when exposed to mechanical treatment such as ball-milling or sonication.[68b, 69] Density gradient ultracentrifugation was used to analyze the number of graphene layers of sonicated samples (**Figure 11**). Next to flakes of few layer graphene a certain fraction of real single layer graphene with a somewhat increased defect density was identified.[70]

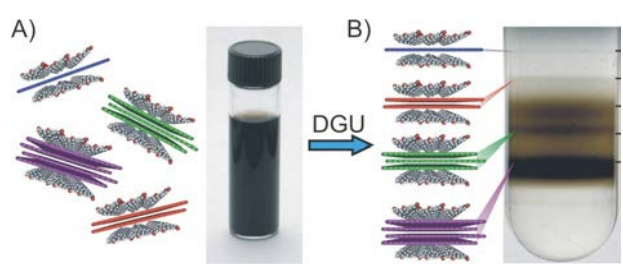

**Figure 11.** A) Polydisperse dispersion of graphene and few-layer graphene stabilized with sodium cholate as surfactant; B) Fractions of graphene, bi-layer and few-layer graphene after density gradient ultracentrifugation. Reproduced with permission.[70] Copyright 2010, American Chemical Society.

In donor-GICs (graphite intercalation compounds) the negatively charged graphene layers, called graphenides, are separated from each other, e. g. by potassium or lithium ions.[47a, 47b, 47d, 71] However, the wet chemical delamination to single layers of graphenide was demonstrated only for flakes with a diameter of about 150 nm.[72] The number of layers can e. g. be counted by the number of frings from HRTEM images.[73]

Acceptor-GICs such as graphite sulfate, can be prepared on the technical-scale and exfoliation can be achieved by inducing thermal decomposition of the intercalated species.[74] Few-layer graphene that partially re-aggregate in the solid are generally obtained by this method.[75] Furthermore, graphene and few-layer graphene can be generated in dispersion directly from an acceptor-GIC using e. g. oleyl amine for stabilization.[76]

GO can be reduced to graphite in solids or in solution and without a stabilizer solids are formed due to aggregation (**Figure 10D**).[77] Here, the defect density depends on the preparation conditions and during thermal reduction additional defects are obviously formed due to carbon loss. GO can also be reduced in dispersion in the presence of a surfactant to form stabilized graphene.[77b, 77c] However, surfactants remain generally strongly adsorbed, although the sodium salt of binol was reported to be removable.[78]

Recently, an efficient electrochemical exfoliation method of graphite was demonstrated, to yield graphene, predominantly bilayer graphene and few-layer graphene in diluted sulfuric acid as reactive solvent and the defect density of bilayer graphene can be estimated to about 0.009%.[79]

Reliable investigations on the functionalization of graphene require graphene with a defect density below 0.5% and e. g. graphene derived from ai-GO fulfills this demand.[29, 52] At a higher defect density changes within the degree of functionalization cannot be detected by Raman spectroscopy, which is the method of choice for the characterization of functionalized samples.

## 4. Non-covalent and Covalent Graphene Chemistry

The functionalization of graphene and few-layer graphene has recently been summarized in some specialized reviews.[6a, 80] Here, we show examples that clearly relate to the functionalization and isolation of functionalized single layers of graphene ($G_1$). Chemical functionalization approaches that lead to functionalized few-layer graphene ($G_{<10}$) or graphite are only briefly mentioned.

In general, non-covalent chemistry is attractive because of the preservation of the conjugated π-system. The non-covalent functionalization is based on weak interactions between graphene and a binding partner e. g. a surfactant which can also be considered as a ligand. Graphene derived from GO was also combined with surfactants for stabilization.[81]

For the covalent functionalization of graphene a covalent bond must be formed what is accompanied with the rehybridization of C-atoms from $sp^2$ to $sp^3$. While C-O bonds are formed during the synthesis of GO, C-C bonds can be formed e. g. using diazonium chemistry, which will be highlighted below.

### 4.1. Non-covalent Approaches

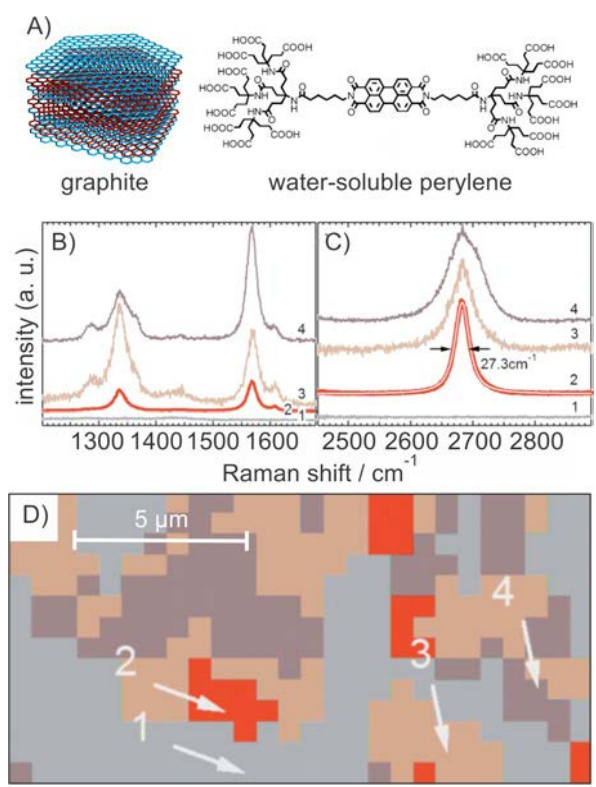

**Figure 12.** A) A water-soluble perylene which is able to exfoliate graphite; B) Raman spectra of delaminated graphene from positions 1-4 in D, showing the D and G peaks, position 2 relates to graphene (G1); C) Raman spectra showing the 2D peaks; $\Gamma 2D < 30$ cm-1 relates to graphene; D) Raman microscopic image coded according to $\Gamma 2D$; substrate (1), graphene (2, $\Gamma 2D$ = 25-39 cm-1), few-layer (3, $\Gamma 2D$ = 39-65 cm-1) and other areas (4, $\Gamma 2D$ > 65 cm-1). Reproduced with permission.[17] Copyright 2009, Wiley-VCH Verlag GmbH & Co.

As depicted in **Figure 11** the interaction of graphite with surface active molecules (surfactants), like sodium cholate,[70a, 82] cetyltrimethylammonium bromide,[83] polyvinylpyrolidone,[84] triphenylene[85] or pyrene derivatives[86] were reported to produce non-covalently functionalized graphene. However, one has to keep in mind that next to single layer graphene $G_1$ also large portions of few-layer graphene and even dispersed graphite are obtained by this approach.

Also coronene carboxylate has been used as surfactant, which allowed for the generation of small flakes of few 100 nm in diameter.

[87] These graphene samples exhibit a defect density in the range of 0.03%. Larger flakes of graphene $G_1$ together with few-layer graphene were obtained using a water-soluble perylene as determined from $\Gamma_{2D}$ in the Raman spectra (**Figure 12**).[17, 88] The water-soluble perylene can delaminate and stabilize graphene with a flake size of about 1 μm with a moderate defect density of approximately 0.01% as indicated by the D peak (**Figure 12**).[17] The presence of defects may be a prerequisite for the successful delamination. The line shape of the 2D peak

clearly indicates the presence of single layer graphene since $\Gamma_{2D}$ is smaller than 39 cm$^{-1}$. The Raman microscopic image (**Figure 12D**) reveals also the polydisperse nature of the sample.

### 4.2. Covalent Approaches

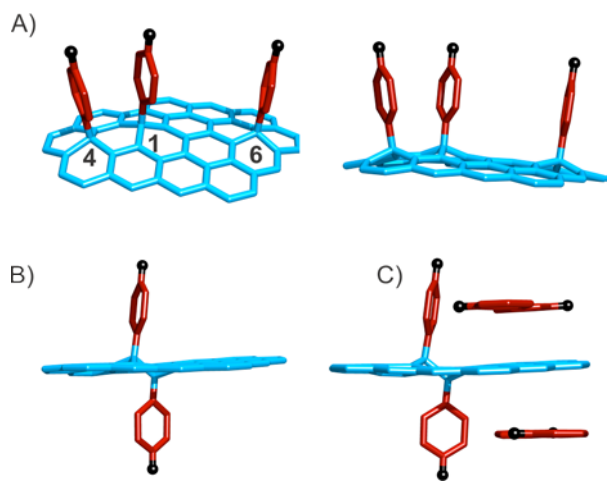

**Figure 13.** A) graphene (blue) functionalized only on the upper side e.g. with aryl moieties (red, black); 1,4- or 1,6-addition patterns are energetically favored and side view: out-of-plane localization of the corresponding sp3-C-atoms;[89] B) side view of graphene functionalized in 1,2-position on both sides of the basal plane; C) an additional non-covalent binding of an aryl moiety by π-π-stacking interactions is shown for comparison.

The covalent chemistry of graphene, few-layer graphene and graphite is a growing field of research and is summarized in several reviews.[80, 90] In principle, wet-chemical functionalization allows for covalent binding to both sides of the graphene plane with a theoretical surface area of 2630 m$^2$/g. However, as illustrated for example in **Figure 13** no exhaustive wet-chemical functionalization of graphene with large organic molecules, such as phenyl groups is possible due to steric reasons, at least when the addends are bound at one side of the basal plane only. Even the complete hydrogenation of graphene, leading to graphane with only sp$^3$-C-atoms has not been realized yet.[91] The highest degree of functionalization approaching the 1:1 stoichiometry was achieved by the reaction of graphene with xenon difluoride to form fluorinated "graphane".[92]

Chemical functionalization of graphene and few-layer graphene in dispersion was investigated using various reactants, including hydrogen, oxygen or halogens, leading to partially functionalized graphene.[91-93] In the following we will show the results of wet- chemical functionalization of graphene on a solid support and the wet chemical functionalization of graphene in dispersion.

*4.2.1. Functionalization of Graphene on a Solid Support*

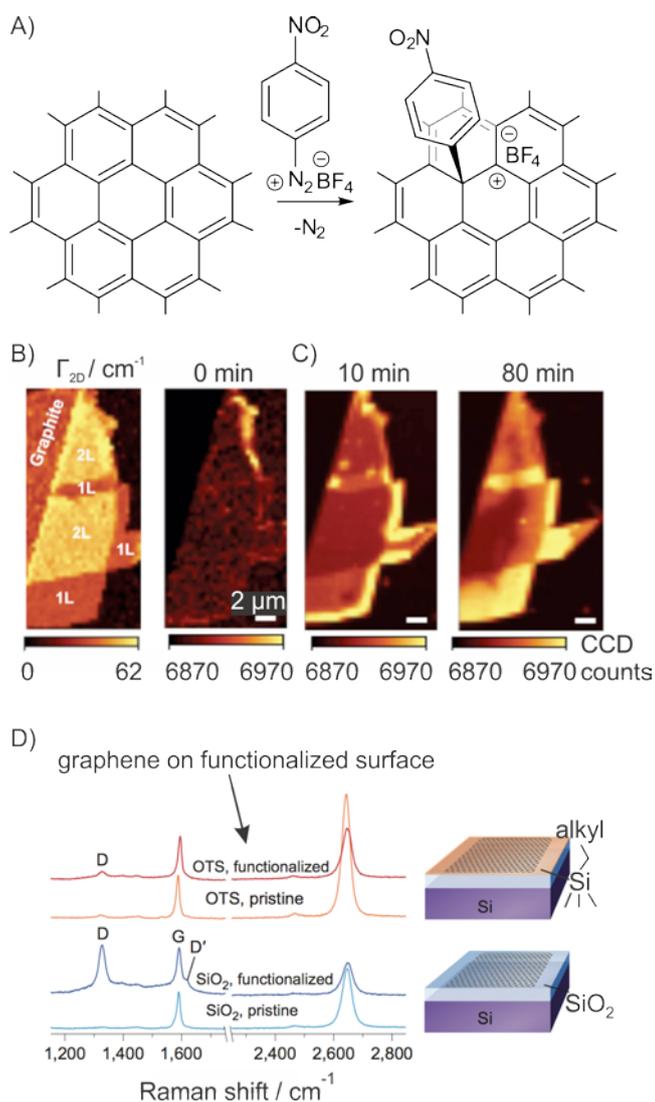

**Figure 14.** A) Reaction of graphene with a diazonium salt; B) Raman micropscopic image of mechanically cleaved graphene, left: mapping of Γ2D and right: D peak intensity; C) mapping of D peak intensity after exposure of graphene to 4-nitrobenzene-diazonium tetrafluoroborate after 10 and 80 min, respectively; Reproduced with permission.[94] Copyright 2010, Wiley-VCH Verlag GmbH & Co. D) Raman spectra of react most likely in *trans*-1,2-position if both sides of graphene are accessible. Next to covalent binding also a competing non-covalent graphene supported on different surfaces before and after functionalization with 4-nitrobenzene-diazonium tetrafluoroborate. Adapted by permission from Macmillan Publishers Ltd: Nature Chemistry,[41] copyright (2012).

In a first series of studies graphene supported on $SiO_2$ was treated with electrophiles to study their reactivity toward graphene.[95] Theoretical calculations suggest that addends favorably add in *cis*-1,4- or *cis*-1,6-position if only one side of graphene is accessible for reactants (**Figure 13**).[89] In contrast to that, addends react most likely in *trans*-1,2-position if both sides of graphene are accessible. Next to covalent binding also a competing non-covalent adsorption of reactants has to be considered, when reaction products are characterized.

A comparatively intensively investigated reaction type is the reaction of aryl diazonium compounds with graphene.[41, 96] **Figure 14** presents SRM images obtained after the treatment of graphene supported on SiO$_2$ with 4-nitrobenzene-diazonium tetrafluoroborate. The reaction most likely involves an electron transfer from graphene to the diazonium ion followed by extrusion of N$_2$ and a subsequent addition of the aryl radical to the oxidized graphene layer. However, further investigations are required in order to understand all details of the conversion. Using SRM the degree of functionalization of edges, central parts and bi-layer graphene is visualized by analyzing the D peak intensity or $\Gamma_{2D}$.[94, 97] The analyses reveal that edges of graphene are more reactive than the interior parts of the basal plane of graphene and that graphene is more reactive than bi-layer graphene. The reason for the higher reactivity of graphene may be due to the corrugation of graphene on the surface which is more pronounced for single layers than for bi-layers of graphene. Furthermore, adsorbed diazonium species could be identified in this study as well, as illustrated in **Figure 13C**.

In another approach graphene was deposited on either SiO$_2$ or on an alkyl-functionalized SiO$_2$-surface **Figure 14D**.[41] After the reaction of the diazonium compound Raman spectra reveal the distinct higher reactivity of graphene on SiO$_2$ compared to the alkyl-terminated surface. These approaches demonstrate that neutral graphene is not highly reactive towards diazonium compounds but additional activation can facilitate the conversion.

It is interesting to note that the $I_D/I_G$ ratio of one (**Figure 14D**) indicates a degree of functionalization of about 0.01% and consequently the very small D peak measured after functionalization for graphene placed on the alkylated surface indicates that almost no reaction occured. Thus, activation of graphene can enhance its reactivity, as it was also demonstrated for graphene placed on nanoparticles whereby graphene becomes locally curved.[98]

The wet-chemical functionalization of graphene supported on SiO$_2$ was also carried out in a two step process using reduced graphene (graphenide) as starting material. With graphenide conceptually no oxidized graphene layers have to be generated (see also **Figure 14**) and at the same time they are better reducing agents than neutral graphene itself. First, the supported graphene was reduced by the treatment with a sodium/potassium alloy in dimethoxyethane (DME).[60a] The resulting surface supported graphenide was then reacted with phenyliodide. In this case an electron transfer from graphenide to phenyliodide takes place to form iodide and phenylradicals. The latter add to graphene (**Figure 15**). Other reactions, e. g. photoinduced reactions of graphene with benzoylperoxide were also reported.[99]

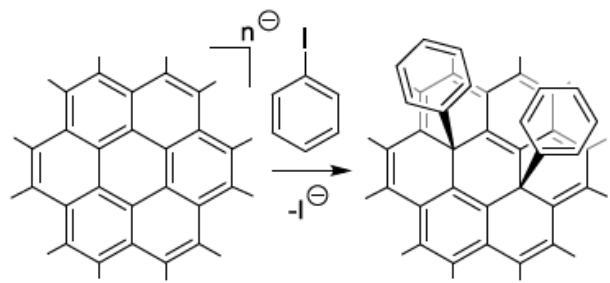

**Figure 15.** Wet chemical reaction of graphenide (activated graphene on a solid support) with phenyl iodide to phenyl-functionalized graphene.[60a]

*4.2.2. Wet-chemical functionalization of graphene in homogeneous dispersion*

Since it has so far not been possible to generate a dispersion of completely exfoliated single layer graphene, chemical reactions are carried out in mixtures including few-layer graphene and dispersed graphite with diameters below 1 µm as predominant species. General approaches for the functionalization have been summarized in the literature.[90d, 100]

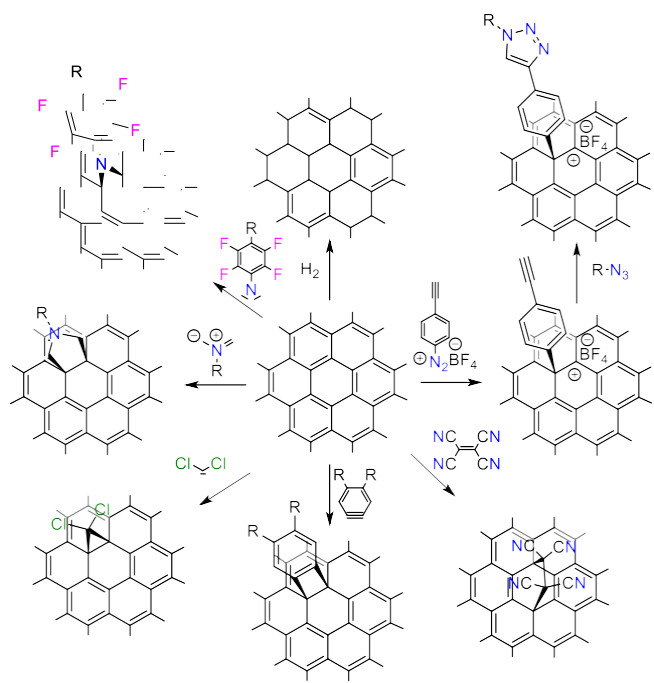

**Figure 16.** Illustration of selected reaction types for the functionalization of graphene and few-layer graphene.

In **Figure 16** typical types of reactions are illustrated, such as hydrogenation,[93c, 93d, 93i] addition of phenylradicals,[101] addition of diazonium species or combined with [3+2]-cycloaddition reactions forming 1,2,3-triazoles.[102] Furthermore, the addition of azomethine ylides,[103] fluorinated phenylnitrene species,[90e] arine species generated from aryl trimethylsilyl triflates,[104] carbenes[105] or Diels- Alder reactions with e. g. tetracyano ethylene were reported.[106] Moreover, acylation reactions were demonstrated to proceed at edges of few-layer graphene.[107] These types of reactions were also applied in order to introduce functional molecules on graphene for generating new properties, e. g. formation of dispersions,[90e, 104] band-gap tuning or light harvesting,[100, 108] and hydrogen storage.[109] Nevertheless, using functionalized graphene in a systematic way for specific applications was not yet conducted. Often, either the single layer nature of functionalized materials is not proven or few-layer graphene is functionalized what leads to covalently functionalized few-layer graphene that can be isolated in stacks as illustrated in **Figure 17A**. To overcome this obstacle for the functionalization of graphene, activation of graphite prior to exfoliation provides an opportunity to address single layers to really synthesize functionalized $G_1$ graphene as illustrated in **Figure 17B**, even if stacks of functionalized $G_1$ graphene ($G_1$-$R_n$) are isolated.

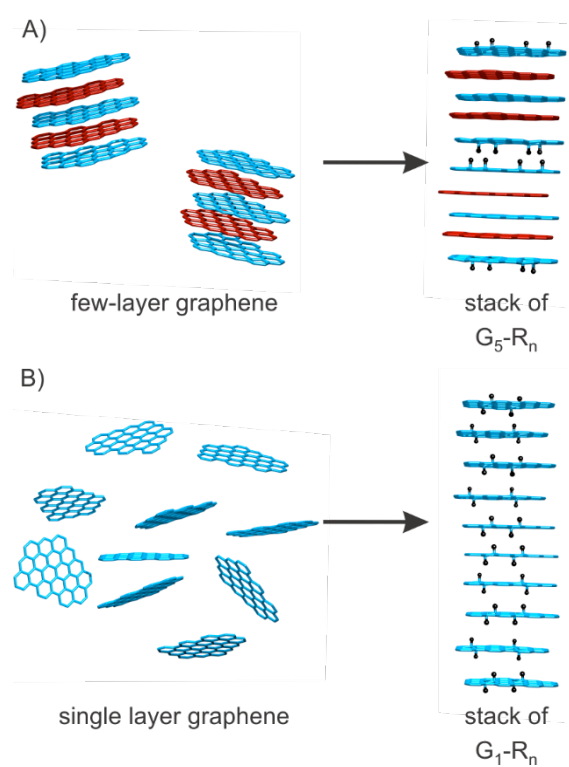

**Figure 17**. Schematic illustrations of the functionalization of A) few- layer graphene as illustrated with five-layer graphene (G5) as it can be generated by dispersing graphite in solvents and the isolated solid functionalization product G5-Rn functionalized with n R groups; B) G1 obtained for example by dispersing alkali metal GICs and its subsequent functionalization to give after work-up stacks of functionalized G1-Rn.

We have recently introduced a very suitable functionalization method for graphene, where negatively charged graphenides were used as activated intermediates for the functionalization in homogenous dispersion. Graphenides are present in donor-GICs where for example alkaline metals serve as electron donors and at the same time as intercalants.[47a, 47b, 47d] Stable examples are $C_6Li$ and $C_8K$, which represent activated reduced graphite. Furthermore, $C_2Li$ is known but can only be formed under high pressure.[110] If donor-GICs are dispersed in a solvent such as DME subsequent addition reactions with electrophiles can be carried out.[111]

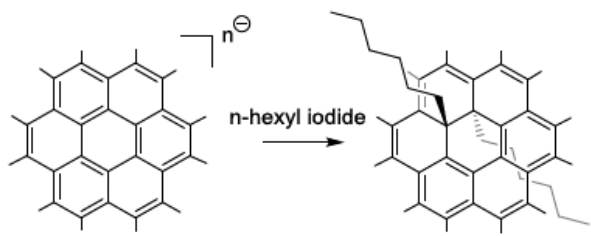

**Figure 18.** Conversion of graphenide to hexyl-functionalized graphene G1-hexyln.[112]

An example is the reaction of graphenide with 4-*tert*- butylphenyldiazonium tetrafluoroborate in DME leading to the formation of arylated graphene $G_1$-aryl$_n$.[63] A similar reaction with *n*-hexyl iodide was demonstrated as well (**Figure 18**).[112] The Raman spectroscopic analysis of a flake of hexylated graphene reveals clear evidence for the single layer nature of functionalized graphene and displays $\Gamma_{2D}$ values < 40 cm$^{-1}$ and $I_D/I_G$ values of about 2 (compare **Figure 8B**). In this example the degree of functionalization varies even within one flake of graphene as determined by SRM.

Although the analytical tools for product characterization improved recently a detailed structural analysis of covalently functionalized graphene remains challenging. For example, it is still not straightforward to distinguish quantitatively between adsorbed and chemically bound species representing a crucial prerequisite to reveal structure property relationships.[63, 113]

## *5. Functionalization of Graphene Oxide*

GO is produced under harsh oxidative conditions and contains oxygen-based addends on both sides of the basal plane, as outlined above. However, synthetic procedures and work-up conditions strongly influence the composition of functional groups.

## 5.1. Degradation of Graphene Oxide

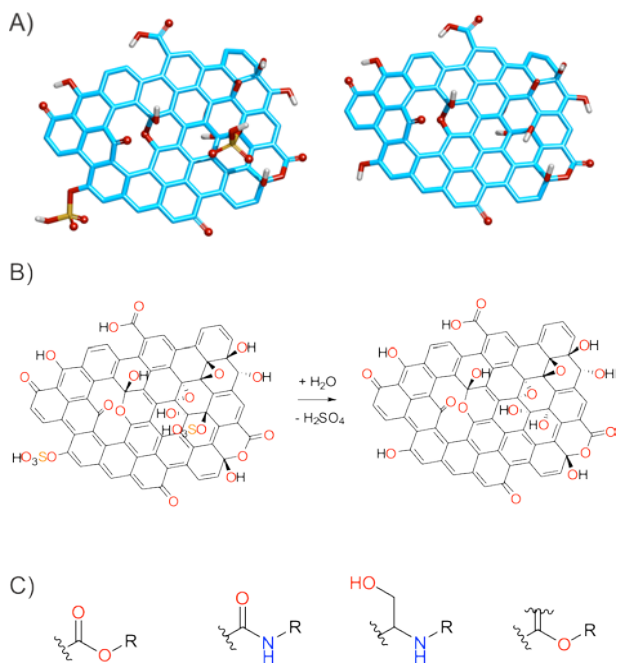

**Figure 19.** A) GO with different functionalities at the basal plane, left: with hydroxyl-, epoxy- and organosulfate groups, right: hydroxyl-, and epoxy groups; B) chemical sketch to illustrate functional groups with a proposed structural defect and the hydrolytical cleavage of organosulfate; C) typical chemical bonds formed for functionalization of GO.

Even at room temperature the binding of functional groups in GO was found to be metastable,[114] and thermally induced $CO_2$ formation can be detected starting at 50 °C.[115] In addition $^{18}O$ from adsorbed $^{18}OH_2$ is incorporated in the cleaved $CO_2$, which is very likely due to the formation of hydrates from carbonyl groups of GO.[115] Furthermore, degradation of GO can be used to partially explain the acidity of GO in water.[116] In steamed GO a large amount of holes was found that lead to porous materials.[117] Moreover, porous graphene was obtained after activating GO with potassium hydroxide before thermal exfoliation.[118] Finally, after prolonged degradation, GO turns into a material that is related to humic acid, as already described by Staudenmaier in 1899.[44]

## 5.2. Transformation of Functional Groups in Oxo- functionalized Graphene

### 5.2.1. Addressing the surface of GO

When parts of the surface of GO are inaccessible for reactants because of coverage with attached substrates the degree of GO- functionalization is limited. Thus, the full potential and efficiency of a reaction is not tapped. Since both sides of GO are highly functionalized the complete delamination has to be be achieved in order to allow for efficient chemical reactions. Density gradient ultracentrifugation studies used to separate GO sheets by size also revealed that some few-layered GO remains present in minor amounts even after sonication.[51b] Furthermore, concentration dependant titrations of GO dispersions with methylene blue reveal that the maximum surface area of GO in water is accessible only at concentrations below 35 µg/ml (**Figure 20**).[119] These experiments suggest that the delamination efficiency should be taken into account for the interpretation of analytical data.

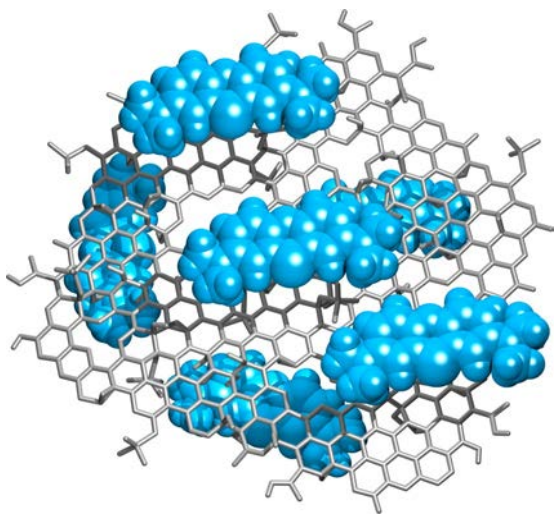

**Figure 20.** A) Illustration of GO (grey) with adsorbed methylene blue (blue) for the determination of the accessible surface area; the maximum surface area of GO is accessible for c(GO) < 35 µg/ml.[119

*5.2.2 Approaches for the functionalization of GO*

Carbonyl or carboxyl groups formed during oxidative graphite degradation can be used for functionalization reactions.[120] In most approaches carboxyl groups are transformed to active esters and subsequently used for the conversion to esters or amides (**Figure 19C**).[121] Furthermore, the direct reaction of amines with graphite oxide and GO was performed as well, leading to partially reduced and functionalized material.[122] The results are summarized in recent reviews.[10a, 123]

Furthermore, several highly porous networks were prepared based on GO or their reduced forms including the formation of aerogels.[124] One example that utilizes hydroxyl groups of GO is the cross-linking of GO sheets with benzene-1,4-diboronic acid forming boronic esters giving a 3D porous network, attractive for gas adsorption (**Figure 21**).[125] Furthermore, organic isocyanates were used for functionalization.[126]

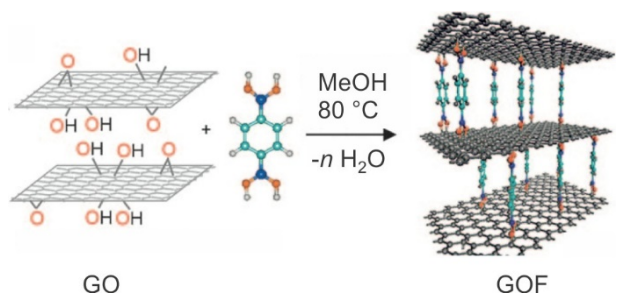

**Figure 21.** Reaction of hydroxyl groups in GO with benzene-1,4- diboronic acid, forming a stable porous framework. Reproduced with permission.[125] Copyright 2010, Wiley-VCH Verlag GmbH & Co.

Due to the amorphous and heterogeneous structure of GO, determination of the amount of different functional groups and the evaluation of the efficiency of chemical reactions are still difficult tasks. Reaction protocols, well known from organic chemistry, are applied on GO and the successful reaction is often evaluated e. g. by dispersibility or performance of the materials in applications.

Nevertheless, it has been reported that GO and its derivatives were used for various applications. Graphene derived from GO was used in transparent electrodes to make touch screens.[127] It was also found that GO can act as a surfactant to disperse carbon nanotubes.[128] Nano-GO with lateral dimensions < 50 nm was functionalized with polyethylenglycol anchored by an amine for drug delivery,[120] and chemo-photothermal therapy.[129] Dye-labeled single strand DNA was non-covalently bound to GO and the fluorescence was found to be quenched due to the interaction of π- systems. Adding a complementary target in nanomolar concentrations restored the fluorescence and this concept was used to detect biomolecules.[130] GO was used in sensors also e. g. to detect humidity with a response speed of about 30 ms only.[131] In addition GO functionalized by organosulfate and $Cs^+$, respectively, were used as hole- and electron-extraction materials in polymer solar cells.[132] Composite materials of GO with small organic molecules or inorganic nanoparticles have been described amongst others e. g. for the preparation of supercapacitors.[6d] For example stearyl amine was used for functionalization of GO to make composite materials with styrene.[133] Moreover, GO and its reduced form were used to make polymer composites applying modern polymerization techniques. [15, 77b, 77c, 134] GO was also found to be a competitive material for charge storage.[135] This listing of the functionalization approaches and applications is far away from being complete. However, in order to optimize functionalization concepts a much more detailed understanding of GO-based chemical reactions is desired, because it remains challenging to determine the local structure of composite materials. Furthermore, it remains difficult to distinguish between functionalization at defect sides, of epoxy groups or others.

*5.2.3 Functionalization of GO at the basal plane*

One approach to more controlled reactions started with the synthesis of ai-GO that bears an almost intact σ-framework of C-atoms.[29] The carbon framework of ai-GO was found to be stable up to 100 °C, even if the functional groups started to cleave or transform already.[136] Furthermore, proof was given that chemical reactions, like the nucleophilic reaction of hydroxide with ai-GO can be applied without degrading the carbon framework by keeping the temperature below 10 °C (**Figure 22A**).[48] In particular, sodium azide was used to substitute organosulfate in ai-GO and to introduce a functional group that is suitable e. g. to develop azide-alkyne cycloaddition reactions in future studies (**Figure 22B**).[31] The degree of functionalization, which can be up to one azide group per 30 C-atoms was determined by the amount of the sulfate leaving group. Furthermore, $^{15}N$ NMR using labeled azide reveals that no adsorbed azide is detectable, proving that covalent bonds have been exclusively formed.

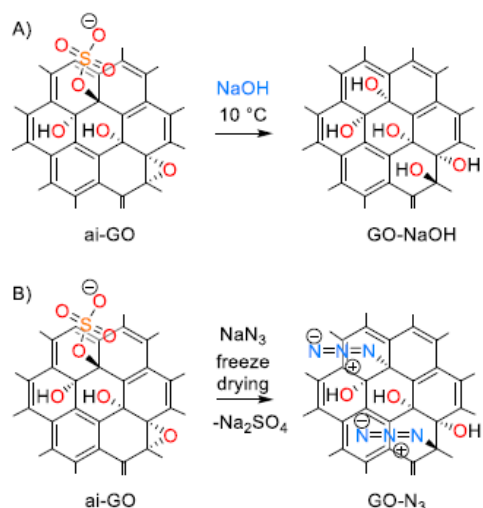

**Figure 22.** A) Reaction of ai-GO with hydroxide without further harming the carbon framework of ai-GO, a prerequisite for controlled chemistry with GO;[48] - Published by The Royal Society of Chemistry; B) nucleophilic substitution of organosulfate groups of ai-GO by azide ions;[31] - Published by The Royal Society of Chemistry.

*5.2.4 Functionalization of reduced GO*

As pointed out above the reduction of normal non ai-GO with a high defect density of approximately 1-3% leads to graphene (also denoted as reduced GO), bearing a substantial amount of □-defects including holes in the basal plane (see also **Figure 4** and **10**). Reduced GO was obtained by reducing GO in water with hydrazine hydrate and used for functionalization reactions as illustrated by a selection of reactions in **Figure 23**. It readily reacts with phenyldiazonium derivatives to provide functional molecules after further derivatization.[108, 137] For the development of polymer composites the diazonium salt of 2-(4-aminophenyl)ethanol was reacted with reduced GO, followed by the reaction with methyl-2-bromopropionate to enable the grafting of styrene by atomic transfer radical polymerization.[138]

Furthermore, the addition of functional groups by carbene chemistry was reported.[139] Water-soluble defective graphene was reported to be formed after partial reduction of GO followed by functionalization with the aryl diazonium salt of sulfanilic acid and a further reduction step.[140] Moreover thermally exfoliated reduced GO was reported to be covalently functionalized by an amine linker with a polymer that reacts with residual epoxy groups at defect sides to form stable dispersions in tetrahydrofuran.[141] Defective graphene was also stabilized by an amphiphilic coil-rod-coil conjugated tri-block copolymer as the stabilizer containing ethylene glycol moieties and acetylene linked phenyl groups.[77b, 77c] This composite is soluble in both organic low polar and water-miscible high polar solvents. Composites of benzylamine reduced GO and citrate stabilized silver nanoparticles were prepared and this composite was found to be efficient for hydrogen peroxide detection.[142] Hydrogen evolution was investigated using nanocomposites of $TiO_2$ and reduced GO as photocatalyst.[143] Furthermore, magnetic nano-composites of reduced GO and $Fe_3O_4$ were also described and are reported to be useful for arsenic removal.[144] Charge storage applications of composites are a popular research field and e. g. composites of reduced GO and $SnO_2$ are reported to perform well.[145] More examples have been recently reviewed.[15, 80a]

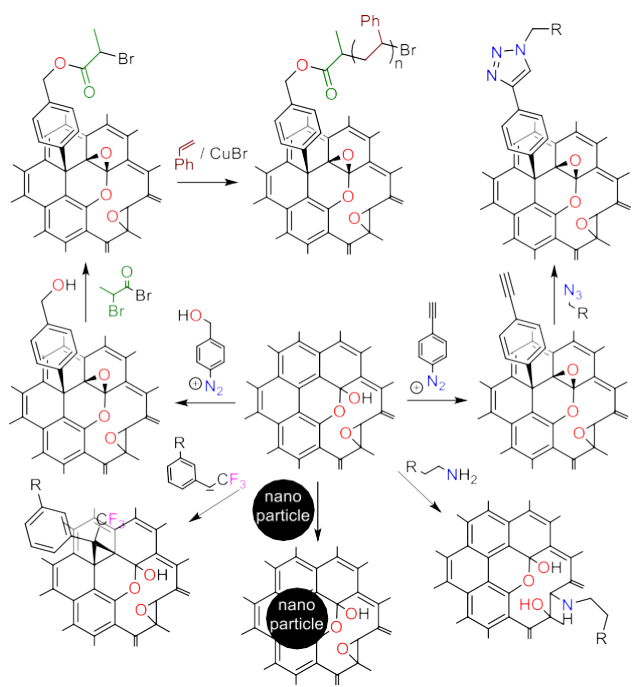

**Figure 23.** Illustration of a selection of reported reactions with reduced GO bearing defects as starting material.

## 6. Conclusions and Outlook

Most approaches for wet chemical graphene functionalization using graphite as starting material have so far led predominantly to the formation of derivatized few-layer graphene ($G_{<10}$-$R_n$) and only a few examples have been published were the formation of truly single layer graphene derivatives $G_1$-$R_n$ could be unambiguously demonstrated. The reason for that is the difficulty in accomplishing quantitative graphite exfoliation before and during the binding of the addends. Nevertheless, a large portion of the surface of graphene layers in graphite can be addressed for the attack of binding partners if graphite is suitably activated and exfoliated prior to the functionalization. This can be accomplished, for example, by using well dispersible ai-GO or negatively charged graphenide as precursors. In the latter case also a pronounced electronic activation of the graphene sheets is guaranteed which allows for extensive redox- and covalent chemistry with electron deficient addends and electrophiles. Following these approaches a series of quite well defined graphene derivatives have recently been published and it can be expected this field will further grow substantially.

The nomenclature of graphene and graphite related compounds that is used in the recent literature is often sloppy and misleading. As a consequence it can be difficult and time consuming to find out what the authors are really talking about. As suggested by Koehler and Stark, a systematic nomenclature for graphene and its derivatives is desirable (**Figure 23**).[95e] We support such a systematic nomenclature approach and propose a general descriptor that is applicable for many types of graphene based systems with different sizes, defect densities, number of layers and degrees of functionalization. Using this description a substrate or adsorbed species can be addressed as well.

$$S/^{s,d}G_n - (R)_f / A$$

**Figure 23.**: S: substrate, s: size of graphene, d: structural defect density of graphene within the carbon framework, G: graphene; n: number of layers of graphene R: addend; f: degree of functionalization; A: non-covalently bound molecules; no S: reactions applied in dispersion.[95e

Applying this scheme on ai-GO, a more precise descriptor would be $5\mu m,0.12\%G_1$-$[(OH)_x(O)_y(OSO_3H)_z]_{50\%}/(H_2O)_{8\%}$ and means that flakes of graphene of 1 layer and a flake size of 5 μm in average and a defect density of about 0.12% in average is functionalized on both sides with an arbitrary ratio of hydroxyl, epoxy and organosulfate groups. There is about one functional group on two carbon atoms and 8 mass-% of water are adsorbed. Few-layer graphene with a size of 150 nm in average can be termed as $^{150\ nm}G_{2-9}$.

Although the concepts for functionalizing graphene lined out in this review are promising a lot of challenges and unsolved problems still remain. Next to the control of the size of flakes used for functionalization, the discrimination between graphene, few-layer graphene and graphite remains difficult to control. These issues have to be addressed in future investigations also in order to establish reliable structure-property relationships. Another important point to address is the qualitative and quantitative determination of defects within samples of graphene, few-layer graphene and GO. Even graphene obtained by CVD methods is not necessarily free from structural defects and we want to point out that Raman spectroscopy alone is not sufficient to prove perfectness of graphene since there are defects known that do not activate the D-peak, as shown for zig- zag edges.[61] Until now it is not fully understood to what extent silent defects activate graphene to enable chemical functionalization.

With GO, the determination of the chemical structure is even more complex since the quantification of different oxygen addends and functional groups remains difficult. Therefore, it is not yet possible to directly determine the defect density in GO and a back conversion to reduced GO is still required to get access to this information. Furthermore, the chemistry of ai-GO with a very low amount of impossible-to-heal σ-defects has just started to emerge. The quantification of functional groups of GO is often determined by methods that are surface sensitive but the bonding state of adsorbed impurities or reagents in many cases cannot be determined quantitatively. Therefore, new analytical approaches must be developed to qualitatively and quantitatively investigate the degree and type of functionalization in the bulk. In the last few years successful functionalization concepts for graphene and GO have been developed and there is no doubt that graphene can indeed be chemically converted to a large extend. In addition GO can be functionalized without degradation of the σ- framework, however, reaction conditions must be well controlled.

At the current level of development it is not clear in detail how the binding structure of chemically functionalized graphene affects its properties in applications. Impurities in graphene derived compounds can play an important role, however, the exact influence is not well addressed until now. As an example, the "metal-free" oxygen reduction using heteroatom doped graphene can be caused by metal impurities.[146] In order to control the physical properties and to enhance the performance of graphene derivatives further fundamental investigations on $G_1$-derivatives are necessary. Knowledge obtained from the chemistry that was successfully performed on other synthetic carbon allotropes such as fullerenes and carbon nanotubes may be a good guide to further improve the functionalization concepts of graphene. Only recently, the controlled synthesis of carbon nanotube derivatives by avoiding side-reactions has been demonstrated.[147] Unwanted side-reactions can even dominate the functionalization of graphene and with respect to that analytical data should be critically discussed. Moreover, the determination of the local structure of functionalized carbon allotropes remains a challenge, and thus STM and HRTEM methods should be further developed. Another possibility to clarify possible chemical structures is using mono-disperse organic model compounds for a given chemical conversion. In this regards, for example, oxygenated aza fullerene derivatives have been studied in detail using NMR spectroscopy and mass spectrometry.[148]

The knowledge generated by systematic graphene functionalization could be a very valuable basis for exploring the chemistry of other sheet materials such as $MoS_2$ or even new, so far unknown synthetic carbon allotropes. One carbon allotrope of interest is graphyne which is composed of sp and $sp^2$ carbon atoms arranged in a 2D- crystal lattice.[149] Finally, applications will benefit from the controlled synthesis of graphene derivatives and the performance of fuel cells, transparent electronics or *in vivo* sensors can certainly improve when defined graphene derivatives will be employed. It can be expected that the full potential of graphene derivatives is not yet exploited but in the future the intensive collaboration of chemists, physicists and material scientists will push the promising technology considerably.